\documentclass[12pt]{article}

\usepackage{amsmath,amsfonts,latexsym,amssymb,amscd}
\usepackage{pslatex}
\usepackage[latin1]{inputenc}
\usepackage[T1]{fontenc}
\usepackage{pspicture}
\usepackage{verbatim,amsthm,curves,graphics}
\usepackage{mathrsfs}

\usepackage{graphicx}

\newcommand{\G}{{\mathcal G}}
\newcommand{\K}{{\mathcal K}}
\newcommand{\mr}{{\mathbb R}}

\newcommand{\mc}{{\mathbb C}}
\newcommand{\mz}{{\mathbb Z}}

\renewcommand{\H}{{\mathcal H}}
\renewcommand{\v}{{\bf v}}

\begin{document}


\title{Uniqueness theorem for 5-dimensional black holes with two
  axial Killing fields}

\author{
Stefan Hollands$^{1}$\thanks{\tt hollands@theorie.physik.uni-goe.de}\:,
Stoytcho Yazadjiev$^{1,2}$\thanks{\tt yazadj@theorie.physik.uni-goe.de}\:,
\\ \\
{\it ${}^{1}$Institut f\"ur Theoretische Physik, 
     Universit\"at G\" ottingen,} \\
{\it D-37077 G\" ottingen, Germany,} \medskip \\ 
{\it ${}^{2}$Department of Theoretical Physics, Faculty of Physics, Sofia
University} \\
{\it 5 J. Bourchier Blvd., Sofia 1164, Bulgaria} \\ 
    }
\date{}

\maketitle

\begin{abstract}
We show that two stationary, asymptotically flat vacuum black holes in 
5 dimensions with two commuting axial symmetries 
are identical if and only if their masses, angular momenta, and 
their ``rod structures'' coincide. We also show that the horizon must 
be topologically either a 3-sphere, a ring, or a Lens-space. 
Our argument is a generalization of 
constructions of Morisawa and Ida (based in turn on key work of Maison) 
who considered the spherical case, 
combined with basic arguments concerning the nature of the factor manifold of 
symmetry orbits. 
\end{abstract}


\sloppy

\section{Introduction}

A key theorem about 4-dimensional stationary asymptotically flat
black holes is that they are uniquely determined by their
conserved asymptotic charges---the mass and angular 
momentum in the vacuum case~\cite{Carter,Robinson}, and 
the mass, angular momentum and charge in the Einstein-Maxwell 
case~\cite{Mazur,Bunting}. 
But the corresponding statement is no longer true in higher
dimensions; there are different vacuum solutions with the same 
mass, and angular momenta~\cite{Myers,Emparan}. 
Nevertheless, it is an
interesting open question whether an analogous statement might still
hold true if a finite number of suitable further parameters associated with
the solution is specified in addition to the mass and angular momenta. 
The purpose of this note is to show that this is indeed true in the 
special case of stationary, asymptotically flat vacuum black holes 
in 5 dimensions which have 2 commuting axial\footnote{By
  this we mean Killing fields whose orbits are periodic. In higher
  dimensions, the set of fixed points of such a symmetry is actually 
generically a higher-dimensional ``plane'', rather than an ``axis.'' We
nevertheless refer to the symmetries as axial, by analogy to the
4-dimensional case.} symmetries with the property that the exterior of the 
spacetime contains no points whose isotropy group is discrete. 
All exact solutions found so far fall in this class.

In fact, what we will show is that the solution is uniquely determined
in terms of its mass, the two angular momenta, and a datum 
called ``rod structure'' that has been introduced in a 
somewhat different form from a more local perspective 
by Harmark~\cite{Harmark,Oelsen}, see~\cite{Reall} for 
a special case\footnote{As we will explain, we also obtain 
new constraints on the rod structure that were not obtained 
in~\cite{Harmark,Oelsen}.}. The rod
structure encodes information about the relative position of the 
various axis and the horizon, and gives a measure of their lengths.
Actually, as we also show, the rod structure in particular determines 
the topology of the horizon, which we show may be either be a 3-sphere $S^3$,
a ring $S^2 \times S^1$, or a Lens-space $L(p,q)$. 
Our proof of these statements uses a known $\sigma$-model 
formulation of the reduced Einstein
equations in 5 dimensions due to Maison~\cite{Maison}, which is analogous 
to a formulation previously found by 
Mazur~\cite{Mazur} and used in his uniqueness proof in 4
dimensions. We combine this technique with 
an elementary analysis of the global structure
of the orbit space of the symmetries. Our result generalizes 
a result of~\cite{Ida} for the special case of $S^3$-horizon topology,
which has a particularly simple rod structure.

In 5 dimensions, it is not known whether an arbitrary stationary, 
asymptotically flat vacuum black hole solution will have 
two commuting axial Killing fields as we are assuming. In fact, the higher
dimensional rigidity theorem~\cite{HIW} only guarantees the existence
of one axial Killing field in such spacetimes in addition to
the timelike Killing field. 
In this regard, the situation in 5
dimensions is very different from the analogous situation in 4
dimensions: Here the original rigidity
theorem~\cite{H72,Hawking,Chrusciel,Racz,Moncrief,Friedrich} 
also guarantees the existence of one 
axial Killing field. But this suffices in 4 dimensions to reduce the 
Einstein equation to the 2-dimensional $\sigma$-model
equations~\cite{Mazur}, and this formulation may then be used to prove the 
uniqueness. By contrast, in 5 dimensions, two axial Killing fields are 
required to make the analogous argument. As we have said, however,
only one axial Killing field appears to be generic. 

\medskip
\noindent
Our {\em conventions and notations} follow those of 
Wald's textbook~\cite{Waldbook}.

\section{Stationary vacuum black holes in $n$ dimensions}

Let $(M,g_{ab})$ be an $n$-dimensional, analytic, asymptotically flat, 
stationary black hole spacetime satisfying the vacuum Einstein 
equations $R_{ab} = 0$, where $n \ge 4$. Let $t^a$ be the 
asymptotically timelike Killing field, $\pounds_t g_{ab} = 0$,
which we assume is normalized
so that $\lim \,  g_{ab} t^a t^b = -1$ near infinity. We denote by 
$H = \partial B$ the horizon of the black hole, $B = M \setminus 
I^-({\mathcal J}^+)$, with ${\mathcal J}^\pm$ the null-infinities of
the spacetime, which are of topology $\mr \times \Sigma_\infty$, with 
$\Sigma_\infty$ a compact manifold of dimension $n-2$.\footnote{In 4
  dimensions, $\Sigma_\infty$ may be shown to be an $S^2$ under
  suitably strong additional hypothesis. A discussion of the structure
of null-infinity in higher dimensions is given in~\cite{HI}.} 
We assume that $H$ is non-degenerate
and that the horizon cross section is a compact connected 
manifold of dimension $n-2$. Under these conditions, one
of the following 2 statements is true: (i) If $t^a$ is tangent to the 
null generators of $H$ then the spacetime must be static~\cite{Sudarsky}. 
(ii) If $t^a$ is not tangent to the 
null generators of $H$, then the higher dimensional rigidity
theorem~\cite{HIW} states that there 
exist $N$ additional linear independent, mutually commuting Killing fields 
$\psi_{1}^a, \dots, \psi^a_{N}$, where $N$ is at least equal to 1. 
These Killing fields generate
periodic, commuting flows (with period $2\pi$), and there exists a
linear combination 
\begin{equation}
K^a = t^a + \Omega_{1}^{} \psi_{1}^a + \dots +
\Omega_{N}^{} \psi_{N}^a, \quad \Omega^{}_{i} \in \mr 
\end{equation}
so that the Killing field $K^a$ is tangent and normal to the null
generators of the horizon $H$, and 
\begin{equation}\label{orth}
K_a \psi_{i}^a = 0 \quad \text{on $H$.}
\end{equation}
Thus, in case (ii), the spacetime 
is axisymmetric, with isometry group $\G=\mr \times U(1)^N$. 
From $K^a$, one may define the surface gravity of the black hole by 
$\kappa^2 = \lim_H (\nabla_a f) \nabla^a f/f$, with $f=(\nabla^a K^b)
\nabla_a K_b$ the norm, and it may be shown that $\kappa$ is constant 
on $H$~\cite{Waldbook}. In fact, the non-degeneracy condition implies 
$\kappa > 0$.

In case (i), one can prove that the spacetime is actually unique, 
and in fact isometric to the Schwarzschild
spacetime~\cite{Israel67} when $n=4$, 
for higher dimensions see~\cite{Gibbons}. 
In this paper, we will be concerned with case (ii). 
We restrict attention to the exterior of the black hole,
$I^-({\mathcal J}^+)$, which we shall again denote by $M$ for
simplicity. We assume that the exterior $M$ is globally hyperbolic. 
By the topological censorship theorem~\cite{Woolgar},
the exterior $M$ is a simply connected manifold (with boundary 
$\partial M = H$). To understand better the nature of the 
solutions, it is useful to bring the field equations 
into a form that exploits 
the symmetries of the spacetime. 
For this, one considers first the factor space $\hat M = M/\G$, where
$\G$ is the isometry group of the spacetime generated by the Killing
fields. Since 
the Killing fields $\psi^a_{i}$ in general have zeros, the 
factor space $\hat M = M/\G$ will normally have singularities.  
We will analyze the manifold $\hat M$ in detail in the next section
for the case $n=5, N=2$.

The full Einstein equations $R_{ab}=0$ on $M$ imply a set of coupled 
differential equations for the metric on the open subsets (of
dimension $d=n-N-1$) of the factor space $\hat M$ corresponding to 
points in $M$ that have a trivial isotropy subgroup\footnote{The 
isotropy subgroup of a point $x \in M$ is the subgroup 
$\{g \in \G; \,\, g \cdot x = x\}$.}. 
To understand these equations in a geometrical way, we note that 
the projection $\pi: M \to
M/\G=\hat M$ defines a $\G$-principal fibre 
bundle over these open subsets of $\hat
M$ (we will call the union of these sets the ``interior'' of $\hat
M$). At each point $x$ in a fibre over $\pi(x)$ in the interior of
$\hat M$, we may uniquely decompose the tangent space $T_x M$ into 
a subspace of vectors tangent to the fibres, and a space $H_x$ of vectors
orthogonal to the fibres. Evidently, the distribution of vector spaces 
$H_x$ is invariant under the group $\G$ of symmetries, and hence forms 
a ``horizontal bundle'' in the terminology of principal fibre 
bundles~\cite{Kobayashi}. According to one of the equivalent
definitions of a connection in the theory of principal
fibre bundles~\cite{Kobayashi}, a horizontal bundle 
is equivalent to the 
specification of a $\G$-gauge connection $\hat D_a$ on the factor
space, whose curvature we denote by $\hat F_{ab}$. The horizontal 
bundle gives an
isomorphism $H_x \to T_{\pi(x)} \hat M$ for any $x$, and 
this isomorphism may be used to uniquely construct a smooth 
covariant tensor field 
$\hat t_{ab \dots c}$ on the interior of $\hat M$ 
from any smooth $\G$-invariant 
covariant tensor field $t_{ab \dots c}$
on $M$. 
For example, the metric $g_{ab}$ on $M$ thereby gives rise to 
a metric $\hat g_{ab}$ on $\hat M$. We let $\hat D_a$
act on ordinary tensors $\hat t_{ab \dots c}$ as the connection of 
$\hat g_{ab}$, with Ricci tensor denoted $\hat R_{ab}$. 

By performing the well-known ``Kaluza-Klein'' reduction of the metric 
$g_{ab}$ on $M$, 
we can locally write the Einstein equations as a system of 
equations on the factor space
$\hat M$ in terms of metric $\hat g_{ab}$, the components 
$\hat F_{I \, ab}, I=0,1,\dots,N$ 
of the curvature and the $(N+1) \times (N+1)$ Gram matrix 
field $G_{IJ}$ 
\begin{equation}
G_{IJ} = g_{ab} X^a_I X^b_J, \quad 
X_I^a = 
\begin{cases}
t^a_{} & \text{if $I=0$}, \\
\psi_{i}^a & \text{if $I=i=1, \dots, N$.} 
\end{cases}
\end{equation}
The resulting equations are similar in nature to the ``Einstein-equations''
on $\hat M$ for $\hat g_{ab}$, coupled to the ``Maxwell fields''
$\hat F_{ab}^I$ and 
the ``scalar fields'' $G_{IJ}$, see~\cite{KK1,KK2}.
 We will not write these equations down 
here, as we will not need them in this most general form. 

The equations simplify considerably if the distribution of
horizontal subspaces $H_x$ is locally 
integrable, i.e., locally tangent to a family of 
$(n-N-1)$-dimensional submanifolds. 
In that case, the connection is flat, 
$\hat F_{ab}^I = 0$, and the dimensionally 
reduced equations may be written as
\begin{equation}\label{reduced2}
\hat D^a ( r G^{-1} \hat D_a G)_I^J = 0
\end{equation}
together with 
\begin{equation}\label{reduced3}
\hat R_{ab} = \hat D_a \hat D_b \log r 
-\frac{1}{4} (\hat D_a G^{-1})^{IJ} \hat D_b G_{IJ} \, .
\end{equation}
The equations are well-defined at 
points in the interior of $\hat M$, corresponding to points with 
trivial isotropy subgroup. At such points, the 
matrix $G$ is not singular, i.e., the Gram determinant
\begin{equation}
r^2 = |{\rm det} \, G|
\end{equation}
does not vanish.
Conversely, one may find stationary
axisymmetric solutions to the Einstein equations by solving 
the above equations subject to appropriate boundary conditions on $\hat M$
which ensure that the metric $g_{ab}$ reconstructed from $\hat
g_{ab}$ and $G_{IJ}$ is smooth. 

Taking the trace of the first
equation, one finds that $r$ is a harmonic function on the interior of 
$\hat M$,  
\begin{equation}
\hat D^a \hat D_a r = 0 \, .
\end{equation}
If $\hat M$ has the structure of a manifold with boundary (as we will
prove in the next section for the situation considered in this paper), 
then on the boundary of $\hat M$ we have $r=0$. We may divide the boundary into a (i) a part corresponding to $H$ where $r=0$ by eq.~\eqref{orth}, and (ii)
a part corresponding to various ``axis,'' where $G_{IJ}$ has a null space 
and where consequently 
one or more linear combinations of the axial Killing fields vanish.
For an asymptotically flat spacetime, the 
quantity $r$ must be approximately equal in an asymptotically
Minkowskian coordinate system to the corresponding quantity formed
from $N$ commuting axial Killing fields and  
$\partial/\partial t$ on exact Minkowski spacetime. 
Thus, in the region of $\hat M$ corresponding 
to a neighborhood of infinity of $M$, and away from the axis, $r \to
\infty$. By the maximum principle, $r$ must therefore be in the range 
$0 < r < \infty$ in the interior of $\hat M$. Thus, in this case, 
the fields $(G^{-1})^{IJ}$ are globally defined on the interior of $\hat M$,
and therefore likewise the dimensionally reduced Einstein equations.  

\section{The factor space $\hat M$}

In this section, we analyze in some detail the factor space $\hat M = M/\G$
in the case when the dimension of $M$ is equal to five. To begin, we consider 
a somewhat simpler situtation in which we have a Riemannian 4-manifold $(\Sigma, h_{ab})$ 
with an isometry group $\K = U(1) \times U(1)$, which may be thought of as a spatial slice 
of our spacetime $M$. We denote the elements of the 
isometry group by $k= (e^{i\tau_1}, e^{i\tau_2})$ with $0 \le \tau_1, \tau_2 < 2\pi$, and we 
denote the Killing vector fields generating the action of the respective $U(1)$ factors by 
$\psi_1^a$ respectively $\psi_2^a$, $\pounds_{\psi_1} h_{ab} = 0 = 
\pounds_{\psi_2} h_{ab}$. These vector fields commute, 
\begin{equation}\label{commutator}
0=[\psi_1, \psi_2]^a = \psi_1^b D_b \psi_2^a - \psi_2^b D_b \psi^a_1 \, .
\end{equation}
We denote the action of a symmetry on a point $x$ 
by $k \cdot x$. As is common, we call the set $O_x = \{ k \cdot x \mid k \in \K \}$ the 
orbit of the point $x$, and we call $\K_x = \{ k \in \K \mid k \cdot x = x \}$ the isotropy subgroup. 
As part of our technical assumptions, we assume that the action is such that there are no points with 
a discrete isotropy subgroup. It is elementary to show that if $\psi_1^a, \psi_2^a$ respectively
$\tilde \psi_1^a, \tilde \psi^a_2$ are two pairs of commuting Killing fields generating such an 
action of $\K$, then they must be related by a matrix of integers $n^i_j$, 
\begin{equation}
\tilde \psi_i^a = \sum_{j=1}^2 n_i^j \cdot \psi^a_j\, , \quad
\left( 
\begin{matrix}
n^1_1 & n^2_1 \\
n^1_2 & n^2_2
\end{matrix}
\right) \in GL(2, {\mathbb Z}) \,\,\, \Leftrightarrow 
\,\,\, {\rm det} 
\left( 
\begin{matrix}
n^1_1 & n^2_1 \\
n^1_2 & n^2_2
\end{matrix}
\right)
= \pm 1 \, .
\end{equation}
We denote the Gram matrix of the Killing fields 
by $f_{ij} = h_{ab} \psi_i^a \psi_j^b$. 

The general structure of the orbit space $\hat \Sigma = \Sigma/\K$ can be 
analyzed by elementary means and is described by the following proposition:

\medskip
\noindent
{\bf Proposition 1:}  
The orbit space $\hat \Sigma=\Sigma/\K$ is a 2-dimensional manifold with boundaries and
corners, i.e., a manifold locally modelled over $\mr \times \mr$
(interior points), $\mr_+ \times \mr$ (1-dimensional boundary
segments) and $\mr_+ \times \mr_+$ (corners). Furthermore, for each of  
the 1-dimensional boundary segments, the rank of the Gram matrix $f_{ij}$ 
is precisely 1, and there is a vector $\v = (v^1, v^2)$ 
with integer entries such that $f_{ij} v^j = 0$ for each point of the segment. 
If $\v_i$ respectively $\v_{i+1}$ 
are the vectors associated 
with two adjacent boundary segments meeting in a corner, 
then we must have
\begin{equation}
\left( 
\begin{matrix}
v^1_i & v^1_{i+1} \\
v^2_i & v^2_{i+1}
\end{matrix}
\right) \in GL(2, {\mathbb Z}) 
\quad
\Leftrightarrow 
\quad
{\rm det} \, (\v_i, \v_{i+1}) = \pm 1
\, .
\end{equation}
On the corners, the Gram matrix has rank 0, and in the 
interior it has  rank 2. 
 
\medskip
\noindent

{\em Proof:}
At each point $x \in \Sigma$, let 
$V_x \subset T_x \Sigma$ be the linear span of the Killing fields 
at $x$, which is tangent to $O_x$, the orbit through $x$. Thus, 
the orbit has the same dimension as a manifold as the vector 
space $V_x$. We let $H_x$ be the orthogonal complement of $V_x$. 
Each point $x \in \Sigma$ must be in precisely one of the sets
\begin{enumerate}
\item[0)] $S_0$, the set of all points such that the dimension of $V_x$ is 0. 
\item[1)] $S_1$, the set of all points such that the dimension of $V_x$ is 1.
\item[2)] $S_2$, the set of all points such that the dimension of $V_x$ is 2.
\end{enumerate}
The set $S_2$ is open because it coincides with the set of all points such 
that the smooth function ${\rm det} \, f$ is different from zero, and the set 
$S_0$ is closed because it is the set of all points where the smooth function 
${\rm Tr} \, f$ is zero. Evidently, if a point $x$ is in $S_i$, then the entire orbit
$O_x$ is in $S_i$, too. We will now show how to construct a coordinate 
chart in a neighborhood of each orbit $O_x$ by considering the 
different cases separately. 

\medskip
\noindent
{\bf Case 2:} If $x \in S_2$, then the orbit $O_x$ has dimension 2. 
In that case, the isotropy group of $x$ can be at most discrete. However, 
this cannot be the case by assumption, so the isotropy group is in fact trivial, and 
this also holds for points in a sufficiently small open neighborhood of $O_x$. If 
we now choose a coordinate system $\{y_1, \dots, y_4\}$ in $\Sigma$ near 
$x$ such that $(\partial/\partial y_1)^a$ and $(\partial/\partial y_2)^a$ are transverse 
to $V_x$, then the surface of constant $y_1=0=y_2$ meets each orbit precisely 
once sufficiently near $x$. Thus, $\{y_3, y_4\}$ furnish the desired coordinate system 
of $\hat \Sigma$ near $x$, showing that this space can be locally modelled over $\mr \times \mr$
near $x$. 

\medskip
\noindent
{\bf Case 1:} For a point $x \in S_1$, the orbit $O_x$ is one-dimensional, i.e., a loop, and 
there exists a linear combination
\begin{equation}\label{sdef}
s^a = v^1 \psi_1^a + v^2 \psi_2^a 
\end{equation}   
such that $s^a$ vanishes on $O_x$, or equivalently $f_{ij} v^i = 0$ there.
Hence, $k = (e^{iv^1 \tau}, e^{iv^2 \tau}), 0 \le \tau < 2\pi$ is 
in the isotropy subgroup $\K_x$. Since $\K_x$ 
is a closed subgroup of the compact group 
$\K = U(1) \times U(1)$, the ratio $v^1/v^2$ must either be rational or $\K_x = \K$. 
The latter would mean that we are in fact in case~0, so we may chose $v^1, v^2$ to 
be integers with no common divisor. It then follows that both $(e^{2\pi i/v^2}, 1)$ and 
$(1, e^{2\pi i/v^1})$ are in the isotropy subgroup. 
Thus, if we follow the loop $O_x$ by acting with 
$(e^{i\tau}, 1)$ on $x$, then we are back to $x$ 
for the first time after $\tau = 2\pi/u^1$, 
where $u^1$ is an integer with $|u^1| \ge |v^2|$, and 
if we likewise follow the loop by acting with $(1, e^{i\tau})$ on $x$, 
then we are back for the first time after $\tau = 2\pi/u^2$, 
where $|u^2| \ge |v^1|$. 
The same holds for any other point in the orbit $O_x$. 

To show that the orbit space $\hat \Sigma$ can be modelled over $\mr_+ \times \mr$ near 
$O_x$, it is useful to construct a special coordinate system $\{y_1, \dots, y_4\}$ near 
$O_x$. This coordinate system is designed in such a way that the action of $\K$ takes 
a particularly simple form. We let $y_4 = u^1 \tau$ be the parameter along the 
orbit $\tau \mapsto (e^{i\tau}, 1) \cdot x$. 
The coordinates $\{y_1, y_2, y_3\}$ measure 
the geodesic distance from the orbit within a suitable tubular
neighborhood, and are defined as follows. First, we pick 
an orthonormal basis (ONB) $\{\tilde e^a_1, \tilde e^a_2, \tilde e^a_3\}$ of $H_x$
and Lie-drag it along the orbit to an ONB at each $x(\tau), 0 \le \tau < 2\pi/u^1$. 
In general, the ONB will not return to itself after we have gone through $O_x$ once, 
i.e., after $\tau = 2\pi/u^1$, but only after we have gone through it $u^1$-times. Consequently, 
by choosing the ONB at $x$ appropriately, we may assume that 
\begin{equation}\label{rot}
(e^{2\pi i/u^1}, 1) \cdot 
\left(
\begin{matrix}
\tilde e^a_1\\
\tilde e^a_2\\
\tilde e^a_3
\end{matrix}
\right)_x = 
\left(
\begin{matrix}
\cos(2\pi w^1/u^1) & \sin(2\pi w^1/u^1) & 0\\
\sin(2\pi w^1/u^1)  & \cos(2\pi w^1/u^1) & 0\\
0 & 0 & 1
\end{matrix}
\right)
\left(
\begin{matrix}
\tilde e^a_1\\
\tilde e^a_2\\
\tilde e^a_3
\end{matrix}
\right)_x \, \, ,
\end{equation}  
for some integer $w^1$.
In order to obtain an ONB of each $H_{x(\tau)}$ varying smoothly as we 
go around the loop $O_x$ once (incuding at $\tau = 2\pi/u^1$), we define 
\begin{equation}
\left(
\begin{matrix}
e^a_1\\
e^a_2\\
e^a_3
\end{matrix}
\right)_{x(\tau)} = 
\left(
\begin{matrix}
\cos(-w^1 \tau) & \sin(-w^1 \tau) & 0\\
\sin(-w^1 \tau)  & \cos(-w^1\tau) & 0\\
0 & 0 & 1
\end{matrix}
\right)
\left(
\begin{matrix}
\tilde e^a_1\\
\tilde e^a_2\\
\tilde e^a_3
\end{matrix}
\right)_{x(\tau)} 
\end{equation}
i.e., we undo the rotation. We now define a diffeomorphism from the solid tube $B^2 \times B^1 \times S^1$ 
into an open neighborhood of $O_x$ by
\begin{equation}
(y_1, y_2, y_3, y_4) \mapsto {\rm Exp}_{x(\tau)}(y_1 e_1^a + y_2 e_2^a + y_3 e_3^a) \, , 
\end{equation}
where $y_4 = u^1 \tau$ is a periodic coordinate with period $2\pi$, and $y_1, y_2, y_3$ are sufficiently small.
($B^2$ is a small open disk around the origin in $\mr^2$ and $B^1$ a small interval around the origin in $\mr^1$.) 
This diffeomorphism defines the desired coordinates. 
By construction, the action of $(e^{i\tau}, 1)$ is given in these coordinates by
\begin{eqnarray}
y_1 &\mapsto& \cos(w^1 \tau) y_1 + \sin(w^1 \tau) y_2\\
y_2 &\mapsto& \cos(w^1 \tau) y_2 + \sin(w^1 \tau) y_1\\
y_3 &\mapsto& y_3\\
y_4 &\mapsto& y_4 + u^1 \tau \, .
\end{eqnarray}

The action of $(e^{iv^1\tau}, e^{iv^2\tau})$ on these coordinates can be found as follows:
First, the action of $(e^{iv^1\tau}, e^{iv^2\tau})$ leaves each point $x(\tau)$ in the 
orbit $O_x$ invariant, and it also maps each space $H_{x(\tau)}$ to itself. Furthermore, 
since $s^a$ and hence $D_a s^b$ is invariant under the action of $(e^{i\tau}, 1)$, 
it follows that the component matrix of $D_a s^b |_{x(\tau)}$ in the ONB $\{e_1^a, e^a_2, e^a_3\}$
commutes with the matrix in eq.~\eqref{rot}. Thus, it must be a rotation in the plane spanned by $e_1^a|_{x(\tau)}, e^a_2|_{x(\tau)}$. 
Therefore, it follows that the action of $(e^{iv^1\tau}, e^{iv^2\tau})$ is given by
\begin{eqnarray}
y_1 &\mapsto& \cos(N \tau) y_1 + \sin(N \tau) y_2\\
y_2 &\mapsto& \cos(N \tau) y_2 + \sin(N \tau) y_1\\
y_3 &\mapsto& y_3\\
y_4 &\mapsto& y_4 
\end{eqnarray}
for some integer $N$. The action of $(1, e^{i\tau})$ on our coordinates may now be determined in the same way, and 
is given in terms of integers $u^2, w^2$ by 
\begin{eqnarray}
y_1 &\mapsto& \cos(-w^2 \tau) y_1 + \sin(-w^2 \tau) y_2\\
y_2 &\mapsto& \cos(-w^2 \tau) y_2 + \sin(-w^2 \tau) y_1\\
y_3 &\mapsto& y_3\\
y_4 &\mapsto& y_4 - u^2 \tau \, .
\end{eqnarray}
Our arguments so far can be summarized by saying that $\{y_1, \dots, y_4\}$ furnish a 
coordinate system covering a tubular neighborhood of the orbit $O_x$, with $y_4$ a $2\pi$ periodic 
coordinate system going around the loop $O_x$ once. The Killing fields $\psi_1^a, \psi_2^a$ are given 
in terms of these coordinates by 
\begin{equation}
\left(
\begin{matrix}
\psi_1^a\\
\psi_2^a
\end{matrix}
\right)
=
\left(
\begin{matrix}
u^1 & w^1\\
-u^2 & -w^2
\end{matrix}
\right)
\left(
\begin{matrix}
l^a\\
m^a
\end{matrix}
\right) \, , 
\end{equation}
where the vector fields $l^a, m^a$ generate the longitude respectively
the meridian of the tori of constant $R = (y_1^2+y_2^2)^{1/2}$ and constant
$y_3$. They are given in terms of the coordinates by
\begin{equation}\label{xidef}
l^a = \left( \frac{\partial}{\partial y_4} \right)^a \, ,
\quad 
m^a = y_1 \left( \frac{\partial}{\partial y_2} \right)^a - y_2 \left( \frac{\partial}{\partial y_1} \right)^a 
\, .
\end{equation}
By the remarks at the beginning of 
this section, since $m^a$ and $l^a$ locally generate an action of $\K$
which has no points with discrete isotropy group, 
the determinant $u^1 w^2 - u^2 w^1$ must be $\pm 1$. 
In view of the definition of 
$s^a$, eq.~\eqref{sdef}, and the fact that $s^a = N m^a$, it also follows that 
\begin{equation}
u^1 v^1 - u^2 v^2 = 0, \quad v^1 w^1 - v^2 w^2 = N \, .
\end{equation}
The first equation implies that $u^1 = cv^2$ and $u^2 = cv^1$ 
for some $c$. Since the modulus 
of $u^1$ is bigger or equal than that of $v^2$ (and the same with 1
and 2 reversed), we must have $|c| \ge 1$. 
In view of the second equation, this implies that $|N|=|c|=1$, and
hence that $u^1 = v^2, u^2 = v^1$. 

The orbit space may now be determined. We have shown that the orbits 
of $(e^{i\tau}, 1)$ and $(1, e^{i\tau})$ have the structure of 
a Seifert fibration, times an interval for the coordinate $y_3$. 
The fibrations are characterized by the winding numbers $(v^2, w^1)$ 
and $(-v^1, -w^2)$ respectively. Thus, for example the first 
fibration is such that as the $l^a$ generator winds around $v^2$-times, 
the generator $m^a$ winds around $w^1$-times, and similarly for 
the other action. Thus, if we  factor by the action of $(e^{i\tau},
1)$, we locally obtain the space $\mr \times 
(\mr^2/{\mathbb Z}_{v^2})$, where ${\mathbb Z}_p \subset U(1)$ is the
cyclic subgroup of $p$ elements whose action on $\mr^2 \cong \mc$ 
is generated by the phase multiplication $z \mapsto e^{2\pi i/p} z$.
The factor $\mr$ in the Cartesian product 
corresponds to the coordinate $y_3$, while the other factor to the coordinates $y_1, y_2$. 
We next factor by $(1, e^{i\tau})$. Since the only nontrivial part of this action on 
$\mr \times (\mr^2/{\mathbb Z}_{v^2})$ is a rotation in the cone $\mr^2/{\mathbb Z}_{v^2}$, we may 
parametrize the orbits in a neighborhood of $O_x$ by $y_3$ and $R=
(y_1^2 + y_2^2)^{1/2}$. This shows that, 
in case (1), $\hat \Sigma$ locally has the structure $\mr \times \mr_+$. 
On the edge locally defined by $R=0$, we have $v^1 \psi_1^a + v^2
\psi_2^a = 0$.

[If we had first factored by  the action of $(1, e^{i\tau})$, we would have locally obtained the space $\mr \times 
(\mr^2/{\mathbb Z}_{v^1})$. The rest would be analogous.]

\medskip
\noindent
{\bf Case 0:} If $x \in S_0$, then $\psi_1^a = 0 = \psi_2^a$ at the
point $x$, and the linear transformations
$D_a \psi_1^b, D_a \psi_2^b$ in the tangent space $T_x \Sigma$
can be viewed as elements of the Lie-algebra $o(4)$ of $O(4)$, defined 
with respect to the Riemannian metric $h_{ab}$ on $T_x \Sigma$. Taking a 
derivative of eq.~\eqref{commutator} and evaluating at $x$, it follows that these 
linear transformations commute at $x$, 
\begin{equation}
(D_a \psi_1^b) D_b \psi_2^c - (D_a \psi_2^b) D_b \psi_1^c = 0 \quad \text{at $x$.}
\end{equation}
This means that, if we form the self-dual and anti-self-dual parts 
\begin{equation}
D_a \psi_{1 b} \pm \frac{1}{2} \epsilon_{ab}{}^{cd} D_c \psi_{1 d}, 
\quad
D_a \psi_{2 b} \pm \frac{1}{2} \epsilon_{ab}{}^{cd} D_c \psi_{2 d}, 
\end{equation}
then the self-dual part of $D_a \psi_{1 b}$ must be proportional to that of 
$D_a \psi_{2 b}$ at $x$, 
and similarly for the anti-self-dual parts, as one may see 
using the Lie-algebra isomorphism 
between $o(4)$ and $o(3) \times o(3)$ corresponding to 
the decomposition into self-dual and anti-self-dual parts. Now pick 
an orthonormal tetrad $\{e_1^a, e_2^a, e_3^a, e^a_4\}$ at $x$. Then  
basis for the 3-dimensional spaces of self-dual and anti-self-dual
skew 2-tensors on $T_x \Sigma$ are given by $e_{1[a} e_{2 b]} \pm e_{3 [a} e_{4 b]}$, 
$e_{1[a} e_{3 b]} \pm e_{2 [a} e_{4 b]}$ and 
$e_{1[a} e_{4 b]} \pm e_{2 [a} e_{3 b]}$, respectively. Performing an
$O(4)$ rotation of the tetrad corresponds to two independent $O(3)$-rotations 
of the respective basis of self-dual and anti-self-dual tensors, and 
vice versa. It follows that tetrad may be rotated if necessary so that 
the self-dual parts of $D_a \psi_{1 b}$ and $D_a \psi_{2 b}$ are proportional 
$e_{1[a} e_{2 b]} + e_{3 [a} e_{4 b]}$, and the anti-self-dual 
parts are proportional to $e_{1[a} e_{2 b]} - e_{3 [a} e_{4 b]}$.  
Therefore we may write, at $x$,  
\begin{equation}\label{relationD}
\left(
\begin{matrix}
D_a \psi_{1 b}\\
D_a \psi_{2 b}
\end{matrix}
\right)
=
\left(
\begin{matrix}
n^1_1 & n^2_1\\
n^1_2 & n^2_2
\end{matrix}
\right)
\left(
\begin{matrix}
2e_{1[a} e_{2b]} \\
2e_{3[a} e_{4b]} 
\end{matrix}
\right) \,  
\end{equation}
for some matrix $n^i_j$. 
Let us now pick Riemannian normal coordinates $\{y_1, y_2, y_3, y_4\}$ 
centered at $x$ corresponding to our choice of tetrad. Then, since the 
Killing fields $\psi_1^a$ and $\psi_2^a$ are globally 
determined by the tensors $D_a \psi_1^b$ and $D_a \psi^b_2$ at 
the point $x$, it follows from eq.~\eqref{relationD} that
$\psi_i^a = \sum n^j_i \cdot s^a_j$ in an open neighborhood of $x$, where
\begin{equation}
s_1^a = y_1 \left( \frac{\partial}{\partial y_2} \right)^a - y_2 \left( \frac{\partial}{\partial y_1} \right)^a 
\, , \quad
s_2^a = y_3 \left( \frac{\partial}{\partial y_4} \right)^a - y_4 \left( \frac{\partial}{\partial y_3} \right)^a \, .
\end{equation}
Since both sets of Killing fields $s_i^a$ and $\psi_i^a$ have periodic orbits with period 
$2\pi$, both the matrix $n^i_j$ and the matrix $v^i_j = (n^{-1})^i_j$ must be 
integer valued. We now define $R_1 =(y_1^2+y_2^2)^{1/2}, R_2 = 
(y_3^2+y_4^2)^{1/2}$. These quantities 
are clearly invariant under the action of $\K$ and in 1---1 correspondence with the orbits near $O_x$. 
This gives $\hat \Sigma$ the structure of $\mr_+ \times \mr_+$ near
the orbit $O_x$. On the edges locally defined by $R_i=0$, we have 
$v_i^1 \psi^a_1+v^2_i \psi^a_2 = 0$. 

We have now constructed the desired coordinate systems in the above
3 cases, and it can be checked that the 
transition functions are smooth. Thus we have 
shown that $\hat \Sigma$ has the structure of a manifold 
with boundaries and corners. \qed

\medskip

The same technique of proof may be used to analyze the possible 
horizon topologies of stationary, asymptotically flat black hole
spacetimes with an action of $\K = U(1) \times U(1)$ satisfying 
the hypothesis that there are no points with discrete isotropy 
group under $\K$. 

\medskip
\noindent
{\bf Proposition 2:} Under the above hypothesis, each 
connected component of the horizon cross section $\mathcal H$
must be topologically either a ring $S^1 \times S^2$, a sphere $S^3$, or 
a Lens-space $L(p,q)$, with $p,q \in \mz$. 

\medskip
\noindent
{\bf Remark 1:} The Lens-spaces $L(p,q)$ 
(see e.g.~\cite[Paragraph 9.2]{Adams}) are the 
spaces obtained by factoring the unit sphere $S^3$
in $\mc^2$ by the group action $(z_1, z_2)
\mapsto (e^{2\pi i/p} z_1, e^{2\pi i q/p} z_2)$. The fundamental 
group of the Lens space is $\pi_1(L(p,q)) = {\mathbb Z}_p$, and 
$q$ is determined only up to integer multiples of $p$. Since 
a Lens-space is a quotient of the positive constant curvature space
$S^3$ by a group of isometries, it
can carry a metric of everywhere positive scalar
curvature, like the other possible topologies $S^3$ and $S^2 \times
S^1$. Thus, the possible horizon topologies listed in Proposition~2
are of so-called ``positive Yamabe type,'' in accordance 
with a general theorem~\cite{Galloway}. 

\medskip
\noindent
{\em Proof:}
As a result of the 
rigidity theorem~\cite{HIW}, we can find a horizon cross section $\mathcal H$
which is itself a Riemannian manifold with induced metric $q_{ab}$, 
of dimension $3$, invariant under the group $\K = U(1) \times U(1)$ of 
axial symmetries generated by $\psi_1^a$ and $\psi_2^a$. 
By the same arguments as in the proof of Proposition~1, $\H$ divided by $\K$ 
is a 1-dimensional manifold with boundary, i.e., a union of intervals, each of 
which corresponds to a connected component of $\H$. We restrict attention 
to one connected component of $\H$, whose space of orbits is a single 
interval. The end points of the interval correspond to 1-dimensional 
orbits where a linear combination of the axial Killing fields vanishes\footnote{
There cannot be points $x$ in $\H$ where both $\psi^a_1$ and $\psi_2^a$ vanish, since 
$D_a \psi^b_1 $ and $D_a \psi_2^b $ would otherwise be two commuting 
but not linearly dependent infinitesimal $SO(3)$ rotations in the
tangent space of $x$, which is impossible.}. 
We call these orbits $O_{x_1}$ and $O_{x_2}$. They are closed loops. All other points of the 
interval correspond to non-degenerate orbits diffeomorphic to the 2-torus 
$S^1 \times S^1$. At $x_1$, an integer linear combination $m^a_1 = v^1_1 \psi_1^a + v^2_1 \psi_2^a$
vanishes, while at $x_2$, an integer linear combination $m^a_2 =
v^1_2 \psi_1^a + v^2_2 \psi_2^a$
vanishes. As in the proof of Proposition~1, we may introduce a local coordinate 
systems in tubular neighborhoods of $O_{x_1}$ and $O_{x_2}$ such that
each neighborhood is diffeomorphic to a solid tube $S^1 \times
B^2$. We denote the radial coordinates measuring the distance from the 
origin in each of the discs $B^2$ by $R_1$ for the first tubular
neighborhood, and by $R_2$ for the second tubular neighborhood. By
construction, the tori of constant $R_1$ respectively $R_2$ correspond
to 2-dimensional orbits of $\K$, i.e., interior points of the
interval. In fact, $R_1$ and $R_2$ measure the distance of the
inteorior point of the interval to the first respectively second boundary
point. 

If $m_1^a, l^a_1$ are the meridian of a torus of constant $r_1$ in the
first tubular neighborhood (with
the longitude going around the $S^1$-direction in the cartesian product
$S^1 \times B^2$), and $m^a_2, l^a_2$ the corresponding quantities for
the second tubular neighborhood, then as in case~1 in the proof of 
Proposition~1, we have
\begin{equation}\label{twist}
\left(
\begin{matrix}
\psi_1^a\\
\psi_2^a
\end{matrix}
\right)
=
\left(
\begin{matrix}
v^2_1  & w^1_1\\
-v^1_1 & -w^2_1
\end{matrix}
\right)
\left(
\begin{matrix}
l^a_1\\
m^a_1
\end{matrix}
\right) 
= 
\left(
\begin{matrix}
v^2_2  & w^1_2\\
-v^1_2 & -w^2_2
\end{matrix}
\right)
\left(
\begin{matrix}
l^a_2\\
m^a_2
\end{matrix}
\right) 
\, .  
\end{equation}
We must now smoothly join the coordinate systems defining the tubular
neighborhoods of $O_{x_1}$ respectively $O_{x_2}$.  
Each tubular neighborhood is a solid torus
$B^2 \times S^1$. Their boundaries (each diffeomorphic to a torus $S^1 \times 
S^1$) must be glued together in such a way that the orbits of $\psi_1^a$ and 
$\psi_2^a$  match. In order to exploit this fact, we act with the
inverse of the second matrix on eq.~\eqref{twist}, to obtain the
relation $m_1^a = p l^a_2 + q m_2^a$, where
\begin{equation}
q = w^1_2 v^1_1-w^2_2 v^2_1 \, , \quad p = v^1_1 v^2_2
- v^2_1 v^1_2 = {\rm det} \, (\v_1, \v_2) \, .
\end{equation}
This means that, while the meridian goes around the the torus bounding
the first tubular neighborhood once, it goes $p$-times around the
longitude and $q$-times around the meridian of the torus bounding 
the second tubular neighborhood. These solid tubes have to be 
glued together accordingly. 
When $p \neq 0 \neq q$, the manifold thereby obtained is topologically 
a Lens space $L(p,q)$ according to one of the equivalent definitions of this space.
Note that $q$ is defined in terms of the vectors $\v_1, \v_2$ by the
above equation up to an integer multiple $sp$, since the vectors ${\bf
w}_1$ respectively, ${\bf w}_2$ are only defined up to integer
multiples of $\v_1$ respectively $\v_2$ by the condition that 
the matrices in eq.~\eqref{twist} have determinant $\pm 1$. However, 
the Lens $L(p,q)$ and $L(p,q+sp)$ are known to be equivalent, so the 
Lens space is determined uniquely by the pair $(\v_1, \v_2)$.

If $q = 0$ modulo $p\mathbb Z$, then $p=\pm 1$, and vice versa. 
In that case, we may similarly argue as above and 
show that $\H$ is topologically $S^3$. Finally, 
if $p = 0$, then $q= \pm 1$ and vice versa, and we may argue as above to show that 
$\H$ is topologically $S^2 \times S^1$. 
\qed

\medskip
\noindent
{\bf Remark 2:}
The proof shows 
how the different topologies $S^3, S^2 \times S^1, L(p,q)$ 
are related to the kernel of 
the Gram matrix $G_{ij} = g_{ab} \psi^a_i \psi_j^b$ 
at the 2 boundary points of the interval
$I = \H/\K$, i.e., the ``rod-vectors'' introduced in the next 
section: If we denote the integer-valued vectors in the kernel by 
$\v_1, \v_2$, and set $p={\rm det} \, (\v_1, \v_2)$, then the topolgy of $\H$
is $S^2 \times S^1$ if $p=0$, it is $S^3$ if $p= \pm 1$, and a Lens space
$L(p,q)$ otherwise.  

\vspace{1cm}

We finally consider in detail the orbit space 
$\hat M= M/\G$ of a stationary, asymptotically flat, Lorentzian
vacuum 
black hole spacetime $(M,g_{ab})$ of dimension $5$ with $2$-dimensional axial 
symmetry group $\K = U(1) \times U(1)$. 
The Killing field $t^a$ that is timelike near infinity corresponds to the isometry group $\mr$, so that the full symmetry group is
$\G=\K \times \mr$. As above, we assume that there are no points in the exterior of $M$ 
whose isotropy subgroup $\K_x$ is discrete. 
We denote the exterior of the black hole again by $M$, so that $M$ itself 
is a manifold with boundary $\partial M = H$. We also assume 
that $M$ is globally hyperbolic. First, we note that 
$t^a$ can nowhere be equal to a linear combination of the 
axial Killing fields. Indeed, letting $F_\tau$ be the flow of $t^a$, 
if $t^a$ were a linear combination of 
the axial Killing fields at a point $x \in M$, then the $F_\tau$-orbit through
$x$ would either be periodic (for a rational linear combination), or
almost periodic (for an irrational linear combination). This
would imply that there are closed (or nearly closed)
$F_\tau$-orbits. However, consider the intersection $S_\tau$ of
$\partial J^+(F_\tau(x))$ with ${\mathcal J}^+$. Evidently, on the one
hand, $S_\tau$ must be bounded as $\tau$ varies, because the orbits 
$F_\tau$ are periodic, or almost periodic. On the other hand, near
${\mathcal J}^+$, the Killing field $t^a$ is timelike, so the sets
$S_\tau$ are related by a time-translation, and hence cannot be
bounded as $\tau$ varies. Thus $t^a$ cannot be tangent to 
a linear combination of the axial Killing fields at any point.

Next, we show that the linear span $V_x$ of  
$\psi^a_{1}, \psi_2^a$ is everywhere spacelike. Indeed
if there was a linear combination $\xi^a$ of the axial Killing fields 
that was timelike or null somewhere, then we could consider the timelike 
or null orbit of $\xi^a$. This orbit must necessarily have a closure
in $M$ that is non-compact, again invoking the global causal structure 
of $M$. On the other hand, $\xi^a$ is a linear combination of axial
Killing fields, so it must have either periodic or almost periodic 
orbits and its closure must hence be isometric to a compact factor group of 
$\K$, a contradiction.

Thus, we have now learned that $V_x$ is spacelike for all $x$, and 
that $t^a$ is transverse to $V_x$ for all $x$. This can now be used 
to determine the general structure of the orbit space $\hat M$. 
To do this, we split the isometry group $\G=\K\times \mr$ into the 
subgroup $\mr$ generated by $t^a$, and the compact subgroup $\K$ generated by the axial Killing fields. Proceeding as 
in the proof of Proposition~1, we first consider the factor space 
$M/\K$. Using that $V_x$ are everywhere spacelike, it now follows
that $M/\K$ is a 3-dimensional manifold with boundaries and corners
(of dimension 2 and 1 respectively). We then factor in addition 
by the subgroup $\mr$. Since the action of $\mr$ is nowhere tangent to
the orbits of $\K$, the action is free, and we find that 
$\hat M = (M/\K)/\mr$ is a 2-dimensional manifold with boundaries and 
corners. 

Finally, we know that $M$ is simply connected by the 
topological censorship theorem~\cite{Woolgar,Woolgar1}. 
By standard arguments from homotopy theory, because 
$\G$ is connected, also 
the factor space $\hat M$ has to be simply connected. We summarize 
our findings in a Proposition:

\medskip
\noindent
{\bf Proposition 3:} Let $(M,g_{ab})$ be the exterior of 
a stationary, asymptotically flat, 5-dimensional vacuum black hole spacetime 
with isometry group $\G = \K \times \mr$, as described above.
Then the 
orbit space $\hat M=M/\G$ 
is a simply connected, 2-dimensional manifold with boundaries and
corners.  Furthermore, in the
interior, on the 1-dimensional boundary segments (except the 
piece corresponding to $H$), and on the corners, 
the Gram matrix $G_{ij} = g_{ab} \psi_{i}^a \psi^b_{j}$ has rank precisely
$2,1$ respectively $0$.

\section{Classification of 5-dimensional stationary spacetimes}

We now consider again
the reduced Einstein equations for a stationary black hole spacetime
with $n-3$ commuting axial Killing fields. We assume 
that the action isometry group $\K$ generated by the axial symmetries 
is so that there are no points with discrete isotropy group.
We also assume in this section that 
the infinity is metrically and topologically a sphere, 
$\Sigma_\infty = S^{n-2}$. Then $n-3$
commuting axial Killing fields are only possible when $n=4,5$
but not for dimensions $n \ge 6$, because the compact part 
$SO(n-2)$ of the asymptotic symmetry group 
admits at most $(n-2)/2$ mutually commuting generators\footnote{If 
we assume a different topology and metric structure of $\Sigma_\infty$, 
such as $\Sigma_\infty = T^{n-2}$, then the spacetime may have 
$n-2$ commuting axial Killing fields.}. 
When $n=4$, the rigidity 
theorem~\cite{Hawking,Chrusciel,Racz,Friedrich} 
guarantees the existence of at least one more
axial Killing field, so that the total number of Killing fields is at
least 2. Thus, for $n=4$ we are always in the situation just
described. If $n=5$, the higher dimensional rigidity
theorem~\cite{HIW} also guarantees at least one more axial Killing
field, but for a solution with precisely one extra axial Killing field, we would 
not be in the situation just described if such solutions were to exist. 
From now on, we take $n=5$, and we {\em postulate} that 
the number of axial Killing fields is $N=2$. We also assume that the 
axial symmetries have been defined so as to act like 
the standard rotations in the $12$-plane
resp. $34$-plane in the asymptotically Minkowskian region.

As explained in Proposition~3 in 
the last section, in that case the factor space $\hat M$
is a simply connected 2-dimensional manifold with boundaries and
corners. As in 4 dimensions, 
one can show using Einstein's equations and Frobenius'
theorem that the 
horizontal subspaces $H_x$ orthogonal to the Killing fields are
locally integrable~\cite{Waldbook}, so the metric may be written as 
\begin{equation}\label{metric}
g_{ab} = (G^{-1})^{IJ} X_{Ia} X_{Jb} + \pi^* \hat g_{ab}
\end{equation}
away from points where $G$ is singular, 
where $\pi: M \to \hat M = M/\G$ is the projection. Furthermore, 
using that ${\rm det} \, G$ is nowhere vanishing in the 
interior of $\hat M$ and negative near infinity, it follows that 
$\hat g_{ab}$ is a metric of signature $(++)$, i.e., a Riemannian metric.  
The reduced Einstein equations for this metric are given by 
eqs.~\eqref{reduced2} and~\eqref{reduced3}. 

Since $\hat M$ is an (orientable) simply connected 2-dimensional 
analytic manifold with boundaries and corners, 
we may map it analytically to the upper complex 
half plane $\{\zeta \in \mc; \,\, {\rm Im} \, \zeta > 0\}$ by the 
Riemann mapping theorem. Furthermore, since $r$ is harmonic, we can
introduce a harmonic scalar field $z$ conjugate to 
$r$ (i.e., $\hat D^a z = \hat \epsilon^{ab} \hat D_b r$). 
Since an analytic mapping is conformal we also have 
$\partial_\zeta \partial_{\bar \zeta} r = 0 = 
\partial_\zeta \partial_{\bar \zeta} z$, and from this, together with 
the boundary condition $r=0$ for ${\rm Im} \, \zeta = 0$, one can 
argue that $\zeta = z + ir$ by a simple argument involving the maximum
principle~\cite{Weinstein}. In particular, $r$ and $z$ are globally 
defined coordinates, and the metric globally takes the form
\begin{equation}\label{hatg}
\hat g_{ab} = e^{2\nu(r,z)} [(dr)_a (dr)_b + (dz)_a (dz)_b] \, .
\end{equation}
Since~eq.\eqref{reduced2}
is invariant under conformal rescalings 
of $\hat g_{ab}$, and since a 2-dimensional metric is conformally
flat, it decouples from eq.~\eqref{reduced3}. In fact, writing the 
Ricci tensor $\hat R_{ab}$ of~\eqref{hatg} 
in terms of $\nu$, one sees that eq.~\eqref{reduced3}
equation may be used to determine $\nu$ by a simple integration. 

The coordinate scalar fields $r,z$ on $\hat M$ are uniquely defined by
the above procedure up to a global conformal transformation of the
upper half plane, i.e., a fractional transformation of the form 
\begin{equation}
\zeta \mapsto \frac{a\zeta + b}{c\zeta + d}, \quad a,b,c,d \in \mr,
\quad ad-bc = 1 \, , \quad \zeta = z+ir \, .
\end{equation}
We will now show how $r, z$ can in fact be uniquely fixed by 
a suitable condition near infinity, up to a translation of $z$.
For 5-dimensional Minkowski spacetime, the Killing fields
$\psi_{1}^a = (\partial/\partial \phi_1)^a$ and $\psi_{2}^a = 
(\partial/\partial \phi_2)^a$ are rotations in the $12$-plane and
the $34$-plane, and the coordinates $r,z$ as constructed above 
are given in terms of inertial coordinates by 
$r=R_1 R_2$ and $z = \frac{1}{2}(R_1^2-R_2^2)$, with $R_1 =
\sqrt{x_1^2+x^2_2}$ and $R_2 = \sqrt{x_3^2+x^2_4}$, as well 
as $\phi_1 = \arctan (x_1/x_2)$ and $\phi_2 = \arctan (x_3/x_4)$. The conformal 
factor is given by $e^{2\nu} = 1/2\sqrt{r^2+z^2}$. In the general 
case, we may pick an asymptotically Minkowskian coordinate system 
and we may define the quantities $r,z$ on the curved, axisymmetric
spacetime under consideration so that they are approximately equal 
near infinity to the
expressions in Minkowski spacetime as just given. In particular, 
we may achieve that 
\begin{equation}\label{bndy}
e^{2\nu} \to \frac{1}{2\sqrt{z^2 + r^2}}
\end{equation}
near infinity, which corresponds to $r \to \infty$, as $z$ is fixed or
to $z \to \pm \infty$ for $r=0$.  
This condition fixes $a=d=1, c=0$ and hence 
leaves only the freedom of shifting $z$ by a constant. 
Thus, in summary, the Einstein equations are reduced to the two 
decoupled equations~\eqref{reduced2} and~\eqref{reduced3}
on the factor manifold
$\hat M = \{ \zeta = z+ir \in \mc; \,\, {\rm Im} \, \zeta > 0\}$ 
with 
metric~\eqref{hatg} and a preferred coordinate system 
$(r,z)$ that is determined up to a translation of $z$. 
The function $\nu$ is determined by 
eq.~\eqref{reduced3}, subject to the boundary condition~\eqref{bndy}.

So far, our construction is similar to 
well-known constructions leading the the uniqueness theorems 
in $n=4$ spacetime dimensions (for a review, see~\cite{Heusler}). In fact,
the only apparent difference to 4 dimensions is that the matrix field
$G_{IJ}$ is a $3 \times 3$ field in 5 dimensions, while it is a $2
\times 2$ matrix field in 4 dimensions. In particular, all information about 
the topology of $M$ and the horizon might seem to be lost. 
In 4 dimensions, the reduced Einstein 
equations may be used to prove that stationary metrics are unique
for fixed mass and angular momentum. On the other hand, 
it is known that in 5 dimensions, solutions are not uniquely fixed by 
these parameters, and that there are even different possibilities
for the topology of the horizon. Thus, one naturally 
wonders where those differences are encoded in the above formulation.  

To understand this point, we must remember that the 2-dimensional 
orbit space $\hat M$ is a manifold with boundaries and corners by Proposition~3. 
The line segments of the boundary correspond to the axis (i.e., the 
sets where a linear combination $v^1 \psi_{1}^a + v^2 \psi_{2}^a$
vanishes), or to the factor space of the horizon, $\hat H=H/\G$.
The corners---the intersections of the line
segments---correspond to points where the axis intersect (i.e., where
both Killing fields vanish simultaneously), or to points where 
the axis intersect the horizon $H$. 
In the realization of $\hat M$ as the upper complex half plane, the
line segments of $\partial \hat M$ correspond to intervals
\begin{equation}
(-\infty, z_1), (z_1, z_2), \dots, (z_k, z_{k+1}), (z_{k+1}, \infty)
\end{equation}
of the real axis forming the boundary of the upper half
plane. Evidently, if the horizon is connected as we assume, precisely one 
interval $(z_h, z_{h+1})$ 
corresponds to the horizon. The other intervals correspond to
rotation-axis, while the points $z_j$
correspond to the intersection points of the axis, except for the 
boundary points of the interval $(z_h, z_{h+1})$ representing the
horizon. Above, we argued
that the coordinate $z$ is defined in a diffeomorphism 
invariant way in terms of the 
solution up to shifts by a constant. Consequently, the $k$ positive 
real numbers
\begin{equation}
l_1 = z_1 - z_2, \quad l_2 = z_2 - z_3, \quad \dots \quad l_k = z_k-z_{k+1}
\end{equation}
are invariantly defined, i.e., are the same for any pair of isometric 
stationary black hole spacetimes of the type we consider. 
Thus, they may be viewed as 
global parameters (``moduli'') characterizing the given solution in addition to 
the mass $m$ and the two angular momenta $J_1, J_2$. Furthermore, with 
each $l_j$, there is associated a label which is either a vector 
$\v_j = (v^1_j, v^2_j)$ of integers such that 
the linear combination $v^1_j \psi_{1}^a +
v^2_j \psi_{2}^a$ vanishes, 
or $\v_h = (0,0)$ if we are on the horizon. 
The labels corresponding to the ``outmost'' intervals 
$(-\infty, z_1)$ and $(z_{k+1}, \infty)$ must be $(0,1)$ respectively 
$(1,0)$, because this is the case for Minkowski spacetime, and 
we assume that our solutions are asymptotically flat. Also from
Proposition~1, and the Remark~2 following the proof of Proposition~2, we 
have 
\begin{center}
\begin{tabular}{l|c}
${\rm det} \, (\v_j, \v_{j+1}) = \pm 1$ & if $(z_{j-1},z_j)$
and $(z_j, z_{j+1})$ are not the horizon \\
\hline
${\rm det} \, (\v_{h-1}, \v_{h+1}) = p$ & if $(z_h,
  z_{h+1})$ is the horizon \\
\end{tabular}
\end{center}
Moreover, $p=0$ for $\H \cong S^2 \times S^1$, $p=\pm 1$ for $\H \cong
S^3$, and $\H \cong L(p,q)$ is a Lens-space for other values of $p$.  
The numbers $\{l_j\}$ and the
assignment of the labels 
$\{\v_j\}$ are related to 
the ``rod-structure'' of the 
solution, introduced from a more local perspective in~\cite{Harmark}\footnote{
In~\cite{Harmark}, neither the condition that $v^1, v^2$ be integers, 
nor the determinant conditions for adjacent rod vectors and 
their relation to the horizon topology
were obtained. Furthermore, 
his rod vectors have 3 components, rather than 2.}, see
also~\cite{Reall} for a special case. We will therefore simply call the 
data consisting of  $\{l_j\}$ and the assignments $\{\v_j\}$ 
the rod structure as well.

For 4 dimensional black holes, there is only the trivial rod structure 
$(-\infty, z_1), (z_1, z_2), (z_2, \infty)$, 
with the middle interval 
corresponding to the horizon, and the first and third corresponding to 
single axis of rotation of the Killing field. Furthermore, the rod
length $l_1$ may be expressed in terms of the global parameters $m,J$
of the solution. By contrast, in 5
dimensions, the rod structure can be non-trivial, and in fact differs
for the Myers-Perry~\cite{Myers} and Black Ring~\cite{Emparan}
solutions. For these cases, 
the rod structure is summarized in the following table~\cite{Harmark}: 

\begin{center}
\begin{tabular}{|c|c|c|c|}
\hline
               & Rods & Rod Vectors (Labels) & Horizon Topology \\ 
\hline
Myers-Perry BH & $(-\infty, z_1), (z_1, z_2), (z_2, \infty)$ & $(1,0),
(0,0), (0,1)$ & $S^3$ \\ 
\hline
Black Ring & $(-\infty, z_1), (z_1, z_2), (z_2, z_3), 
(z_3, \infty)$ & $(1,0),
(0,0), (1,0), (0,1)$ & $S^2 \times S^1$\\
\hline
Flat Spacetime & $(-\infty, z_1), (z_1, \infty)$ & $(1,0), (0,1)$ & ---\\
\hline
\end{tabular}
\end{center}
The following rod structure would represent a ``Black Lens'' if 
such a solution would exist:
\begin{center}
\begin{tabular}{|c|c|c|c|}
\hline
               & Rods & Rod Vectors (Labels) & Horizon Topology \\ 
\hline
Black Lens & $(-\infty, z_1), (z_1, z_2), (z_2, z_3), 
(z_3, \infty)$ & $(1,0),
(0,0), (1,n), (0,1)$ & $L(n,1)$\\
\hline
\end{tabular}
\end{center}

\noindent
Even for 
a fixed set of of asymptotic charges $m,J_1,J_2$ the invariant lengths 
of the rods $l_1=z_1-z_2, l_2=z_2-z_3$ 
may be different for the different Black Ring solutions, 
corresponding to the fact that there exist non-isometric 
Black Ring solutions with equal asymptotic charges~\cite{Emparan,Reall}. 
On a rod 
labelled ``$(1,0)$'', all components of $G_{1j}=G_{j1}, j=1,2$  
vanish but not the other ones, 
while on a rod labelled ``(0,1)'', all components $G_{2j}=G_{j2}$ vanish. 
The vector $(1,0)$ hence corresponds to a $\partial/\partial \phi_1$-axis, 
while the vector $(0,1)$ corresponds to a $\partial/\partial
\phi_2$-axis. 
Thus, we see that the rod structure 
enters the reduced field equations through the boundary conditions 
imposed upon the matrix field $G_{ij}$. The horizon topology is also 
determined by the rod structure by Proposition~2, see also 
Remark~2 following that proposition. This is how the different
topology and global nature of the solutions in 5 dimensions 
are encoded in the reduced Einstein equations on the upper half plane $\hat M$.

Clearly, since we have argued that the rod
structure is a diffeomorphism invariant datum constructed from the 
given solution, two given
stationary black hole solutions with 2 axial Killing fields cannot be 
isomorphic unless the rod structures and the masses and angular
momenta coincide. The main purpose of this paper is to point out 
the following converse to this statement:

\medskip
\noindent
{\bf Theorem:} Consider two stationary, asymptotically flat, vacuum 
black hole spacetimes of dimension 5, having two commuting axial Killing 
fields that commute also with the time-translation Killing field.
Assume that both solutions have the same rod structure, and the same 
values of the mass $m$ and angular momenta $J_1,J_2$. 
Then they are isometric.

\medskip
\noindent
{\em Proof:}
As in 4 spacetime dimensions, the key step in the argument is 
to put the reduced Einstein equations in a suitable
form. Following~\cite{Ida} (see also~\cite{Maison}), this is done as
follows in 5 dimensions. On $M$, we first define the two twist 1-forms
\begin{eqnarray}
\omega_{1a} &=& \epsilon_{abcde} \psi^b_{1} \psi^c_{2} \nabla^d
\psi^e_{1}\\
\omega_{2a} &=& \epsilon_{abcde} \psi^b_{1} \psi^c_{2} \nabla^d
\psi^e_{2} \, .
\end{eqnarray}
Using the vacuum field equations and standard identities for 
Killing fields~\cite{Waldbook}, one shows that these 1-forms are closed. 
Since the Killing fields commute, the twist forms are invariant under
$\G$, and so we may define corresponding 1-forms $\hat \omega_{1a}$
and $\hat \omega_{2a}$ on the interior of the factor space 
$\hat M = \{ \zeta \in \mc; \,\, {\rm Im} \, \zeta > 0\}$. 
These 1-forms are again closed. Thus, the ``twist potentials''
\begin{equation}\label{lineint}
\chi_i = \int_0^\zeta \hat \omega_{i\, \zeta} d\zeta + \hat \omega_{i
  \, \bar
  \zeta} d\bar \zeta
\end{equation} 
are globally defined on $\hat M$ and independent of the path connecting 
$0$ and $\zeta$. We introduce the $3 \times 3$ matrix field $\Phi$ by
\begin{eqnarray}
\Phi=
\left(
  \begin{array}{ccc}
    ({\rm det} \, f)^{-1} & -({\rm det} \, f)^{-1}\chi_1 & -({\rm det} \, f)^{-1} \chi_2  \\
    -({\rm det} \, f)^{-1} \chi_1 & f_{11} + ({\rm det} \, f)^{-1}\chi_1\chi_1 &  f_{12} +
    ({\rm det} \, f)^{-1}\chi_1\chi_2 \\
    -({\rm det} \, f)^{-1} \chi_2 & f_{21} + ({\rm det} \, f)^{-1}\chi_2\chi_1 &  f_{22} +
    ({\rm det} \, f)^{-1}\chi_2\chi_2 \\
  \end{array}
\right) \, .
\end{eqnarray}
Here $f_{ij}$ is the Gram matrix of the axial Killing fields, 
\begin{equation}
f = 
\left(
\begin{array}{ccc}
G_{11} & G_{12}\\
G_{21} & G_{22}
\end{array}
\right) \, .
\end{equation}
The matrix $\Phi$ satisfies $\Phi^T=\Phi$, $\det \, \Phi=1$, and is positive
semi-definite, meaning that it may be written in the form
$\Phi= S^T S$ for some matrix $S$ of determinant 1. As a consequence of the 
reduced Einstein equations~\eqref{reduced2} and~\eqref{reduced3}, 
it also satisfies the divergence identity
\begin{eqnarray}\label{divergenceid}
\quad \quad 
\hat D^a [r \Phi^{-1} \hat D_a \Phi] =0
\end{eqnarray}
on $\hat M$.

Consider now the exterior of the 
two black hole solutions as in the statement of the 
theorem, denoted $(M,g_{ab})$ and $(\tilde M, \tilde g_{ab})$. 
We denote the corresponding matrices defined as above by $\Phi$ and
$\tilde \Phi$, and we use the same ``tilde'' notation to distinguish 
any other quantities associated with the two solutions. 
Since the orbit spaces of the respective spacetimes 
can both be identified with the upper half-plane as analytic
manifolds, we can identify the respective orbit spaces. Furthermore, 
one can show by reversing the 
constructions of the local analytic coordinate systems 
in the proof of Proposition~1
that the $\G$-manifold $M$ can be uniquely reconstructed 
from the rod structure, i.e., $M$ as a manifold with a $\G$-action
is uniquely determined by the rod structure modulo diffeomorphisms
preserving the action of $\G$. Therefore, since the rod structures 
$\{\tilde l_j, \tilde \v_j\}$ and $\{l_j, \v_j\}$ are the same, 
$M$ and $\tilde M$ 
are isomorphic as manifolds with a $\G$ action, and we may hence 
assume that $M = \tilde M$, and that $\tilde t^a = t^a, 
\tilde \psi_i^a=\psi_i^a$ for $i=1,2$. It follows 
in particular that 
$\tilde g_{ab}$ and $g_{ab}$ may be viewed as being defined on the 
same analytic manifold, $M$, and we may also assume 
that $\tilde r= r$ and $\tilde z = z$. 
Consequently, it is possible to combine the divergence 
identities~\eqref{divergenceid} for the two solutions into a single 
identity on the upper complex half plane. This key 
identity~\cite{Mazur,Maison} is called the ``Mazur identity'' and is given by:
\begin{eqnarray}
{\hat D}_{a}(r {\hat D}^{a} {\rm Tr} \, \Psi )
= r \, {\hat g}^{ab} {\rm Tr} \, \left[ \Delta {\mathcal J}^T_a \tilde 
\Phi \Delta {\mathcal J}^{}_b \Phi^{-1} \right]
\end{eqnarray}
where
\begin{equation}
\Psi= \tilde \Phi \Phi^{-1}-1, \quad 
\Delta {\mathcal J}_a = \tilde \Phi^{-1} \hat D_a \tilde \Phi -
\Phi^{-1} \hat D_a \Phi \, .
\end{equation}
Using now the identities $\Phi = S^T S$ and $\tilde 
\Phi = \tilde S^T \tilde S$, 
the Mazur identity can be presented in the form
\begin{eqnarray}\label{MazurId}
{\hat D}_{a}(r {\hat D}^{a} {\rm Tr} \, \Psi) = r \, 
\hat g^{ab} {\rm Tr} \, \left[N^{T}_a N^{}_b\right]
\end{eqnarray}
where $N_{a} = S^{-1} \Delta {\mathcal J}_{a} \tilde S$.

The key point about the Mazur identity~\eqref{MazurId} 
is that on the left side we have a total divergence, 
while the term on the right hand side is 
non-negative. This structure is now exploited 
by integrating the Mazur identity over $\hat M$. 
Using Gauss' theorem, one finds
\begin{eqnarray}\label{MazurIdint}
\int_{\partial \hat M \cup \infty} r \hat D_a {\rm Tr} \, \Psi \, 
d\hat S^a
= 
\int_{\hat M} r \, \hat g^{ab} {\rm Tr} \, \left[N^{T}_a N^{}_b\right] \,
d\hat V \, , 
\end{eqnarray}
where the integral over the boundary includes an integration 
over the ``boundary at infinity''.
If one can show that the boundary integral on the left side is 
zero, then it follows that $N^a$ vanishes on $\hat M$, and hence that 
$\tilde \Phi^{-1} \hat D_a \tilde \Phi = \Phi^{-1} \hat D_a \Phi$. 
Since this implies that $\Phi^{-1} \tilde \Phi$ is a constant matrix 
on $\hat M$, one concludes that $\Phi = \tilde \Phi$ if this holds true 
at one point of $\hat M$. Using that the Gram matrices 
$\tilde f_{ij}$ and $f_{ij}$ become equal near infinity, and 
using that $\tilde \chi_i$ is equal to $\chi_i$ on the axis (see below), 
one conclues that $\Phi$ is equal to $\tilde \Phi$ on an axis near infinity, 
and hence equal everywhere in $\hat M$. 

This can now be used as follows to show that the spacetimes are isometric. 
First, it immediately follows from $\tilde \Phi = \Phi$ that $\tilde \chi_i = 
\chi_i$ and
that the Gram matrices of the axial Killing fields are identical 
for the two solutions, $\tilde f_{ij} = f_{ij}$. To see that 
the other scalar products between the Killing fields coincide for the 
two solutions, let $\alpha_i = g_{ab} t^a \psi^b_i, \beta = 
g_{ab} t^a t^b$, and define similarly the scalar products 
$\tilde \alpha_i, \tilde \beta$ for the other spacetime. One derives 
the equation
\begin{equation}
\hat D_a [(f^{-1})^{ij} \alpha_j] = r({\rm det} \, f)^{-1} \, \hat \epsilon_a{}^b
\, (f^{-1})^{ij} \hat D_b \chi_j \, . 
\end{equation}
The right side does not depend upon the conformal factor $\nu$, so 
since $\tilde \chi_i = \chi_i$ and $\tilde f_{ij} = f_{ij}$, it also follows that 
$\tilde \alpha_i = \alpha_i$ up to a constant. That constant has to vanish, since 
it vanishes at infinity. Furthermore, from 
\begin{equation}
\beta = (f^{-1})^{ij} \alpha_i \alpha_j - ({\rm det} \, f)^{-1} r^2
\end{equation}
we have $\tilde \beta = \beta$. 
Thus, all scalar products of the Killing 
fields are equal for the two solutions, $\tilde G_{IJ} = G_{IJ}$ on the 
entire upper half plane. Viewing now the reduced Einstein 
equation~\eqref{reduced3} as an equation for $\nu$ 
respectively $\tilde \nu$, one  
concludes from this that $\tilde \nu = \nu$. Thus, summarizing, 
we have shown that if the boundary integral in the integrated 
Mazur identity eq.~\eqref{MazurIdint} could be shown to vanish, then 
$\tilde G_{IJ} = G_{IJ}$, $\tilde r = r$, $\tilde z = z$ and $\tilde \nu = 
\nu$. Since $\tilde t^a = t^a, \tilde \psi^a_i = \psi_i^a$ 
it follows from eqs.~\eqref{metric} and~\eqref{hatg}
that $\tilde g_{ab} = g_{ab}$. This proves that the two spacetimes are 
isometric, proving the theorem.

Thus, to establish the statement of the theorem, 
one needs to prove that the boundary integral in~\eqref{MazurIdint} vanishes. 
For this, one has to analyze the behavior of the integrand
$r {\hat D}_a {\rm Tr} \, \Psi$ near the 
boundary ${\rm Im} \, \zeta = 0$ (i.e., the horizon and the axis)
and near the boundary at infinity, ${\rm Im} \, \zeta  \to \infty$ 
as ${\rm Re} \, \zeta$ is kept fixed (i.e., spatial infinity).
At this stage one has to use again that the asymptotic charges and 
the rod structure of the solutions are assumed to be identical. 
We divide the boundary region into three parts: (1) The axis, (2)
the horizon, and (3) infinity. 
\begin{enumerate}
\item[(1)] On each segment $(z_j, z_{j+1})$ of the 
real line ${\rm Im} \, \zeta = r = 0$ representing an axis, we know
that the null spaces of the Gram matrices $f_{ij}$ and $\tilde f_{ij}$ 
coincide, because we are assuming 
that the rod structures of both solutions are identical. Furthermore, 
from eq.~\eqref{lineint}, and from the fact that $\hat \omega_a^i$
vanishes on any axis by definition, the twist potentials $\chi_i$
are constant on the real line outside of the segment $(z_h,z_{h+1})$
representing the horizon. The difference between the constant
value of $\chi_i$ on the real line left and right to the horizon
segment can be calculated as follows:
\begin{eqnarray*}
\chi_i(r=0,z_h) - \chi_i(r=0,z_{h+1})
&=& 
\int_{z_h}^{z_{h+1}} \hat \omega_{i \zeta} d\zeta + \hat \omega_{i \bar
  \zeta} d\bar \zeta \\
&=& \frac{1}{(2\pi)^2} 
\int_{{\mathcal H}} \nabla_{[a} \psi_{b] \, i} \, dS^{ab} \\
&=& 
\frac{1}{(2\pi)^2} 
\int_{S^3_\infty} \nabla_{[a} \psi_{b] \, i} \, dS^{ab} = 
{\rm const.} \, J_i \, .
\end{eqnarray*}
The first equality follows from the definition of the twist 
potentials, the second from the defining formula for the twist
potentials and the fact that the twist potentials are
invariant under the action of the 2-independent rotation isometries 
each with period $2\pi$ (with ${\mathcal H}$ a horizon cross section), 
the third equation follows from Gauss' theorem and the fact that 
$\nabla^a (\nabla_{[a} \psi_{b] \, i})=0$ when $R_{ab}=0$, and the last
equality follows from the Komar expression for the angular momentum in 
5 dimensions. The analogous expressions hold in the spacetime $(\tilde
M, \tilde g_{ab})$. 
Because we assume that $J_i = \tilde J_i$, we can 
add constants to $\chi_i$, if necessary, so that $\chi_i = \tilde
\chi_i$ on the axis, and in fact that $\Delta \chi_i = \chi_i - \tilde
\chi_i = O(r^2)$ near any axis. 
One may now analyze the contributions to the boundary integral 
coming from the axis using the expression
\begin{equation}\label{difference}
\hat D_a {\rm Tr} \, \Psi
= \hat D_a \Bigg[
({\rm det} \, \tilde f)^{-1}
\{- \Delta ({\rm det} \, f) + (f^{-1})^{ij} \Delta \chi_i \Delta \chi_j\}
+ (f^{-1})^{ij} \Delta f_{ij}
\Bigg] \, .
\end{equation}
We consider a particular axis rod with rod vector $\v=(v^1, v^2)$, 
which by assumption is identical for the two solutions. 
We pick a second basis vector ${\bf w}=(w^1, w^2)$, and we denote by 
$\v^*, {\bf w}^*$ the dual basis. We conclude that, on the given rod
\begin{equation}
f_{ij} = a v_i^* v_j^* + b w_i^* w_j^* + 2c v_{(i}^* w_{j)}^*, \quad  
\tilde f_{ij} = \tilde a v_i^* v_j^* + \tilde b w_i^* w_j^* + 2 \tilde
c v_{(i}^* w_{j)}^*, 
\end{equation}
with $\tilde a = O(r^2) = a, \tilde b = O(1) = b$ and 
$\tilde c = O(r) = c$. We insert this into eq.~\eqref{difference}, 
we use that $\Delta \chi_i = O(r^2)$ on the axis, 
and we use the detailed fall-off properties of the metric 
as well as the quantities $\chi_i, f_{ij}, \tilde \chi_i, \tilde f_{ij}$
for large $z$, which are the same for any asymptotically flat 
solution to the relevant order. One finds that $\hat D_a {\rm Tr} \, \Psi$ is 
finite on the axis, so that the corresponding contribution to the line 
integral vanishes. The details of this analysis are in close 
parallel to the
corresponding analysis of Ida et al.~\cite{Ida}, who analyzed 
the situation for a special horizon topology and rod structure.
\item[(2)]
On the horizon segment, the matrices $f_{ij}, \tilde f_{ij}$ are
invertible, so $\hat D_a {\rm Tr} \, \Psi$ is regular, and the boundary integral over this segment 
vanishes. 
\item[(3)]
Near infinity, one has to use the asymptotic form of the metric 
for an asymptotically flat spacetime in 5 dimensions. In an 
appropriate asymptotically Minkowskian coordinate system
such that asymptotically $\psi_{1}^a = (\partial/\partial \phi_1)^a$
and $\psi_{2}^a = (\partial/\partial \phi_2)^a$, 
it takes the form
\begin{eqnarray*}
g &=& - \Bigg( 1- \frac{\mu}{R^2} + O(R^{-2}) \Bigg) dt^2 + 
\Bigg( 
\frac{2\mu a_1(R^2 + a^2_1)}{R^4} \sin^2 \theta + O(R^{-3})
\Bigg) dt d\phi_1  \\
&& + \Bigg( 
\frac{2\mu a_2(R^2+a^2_2)}{R^4} \cos^2 \theta + O(R^{-3})
\Bigg) dt d\phi_2 + \Bigg( 1 + \frac{\mu}{2R^2} + O(R^{-3}) 
\Bigg) \times \\
&& \times \Bigg( 
\frac{R^2 + a_1^2 \cos^2 \theta + a_2^2 \sin^2 \theta}{
(R^2+a^2_1)(R^2+a_2^2)} R^2  \, dR^2 + (R^2 + a^2_1 \cos^2 \theta 
+ a^2_2 \sin^2 \theta) \, d\theta^2 \\
&& + (R^2 + a^2_1)
\sin^2 \theta d\phi^2_1 + (R^2 +a_2^2) \cos^2 \theta d\phi^2_1 
\Bigg)
\end{eqnarray*}
where $\mu,a_1,a_2$ are parameters proportional to the mass, and the 
angular momenta $J_1, J_2$ of the solution. One must then 
determine the functions $z,r$ as functions of $R,\theta$
near infinity using the reduced Einstein equations, subject to the 
boundary condition~\eqref{bndy} near infinity. This then 
enables one to find asymptotic expansions for $f_{ij}, \tilde f_{ij}, 
\chi_i, \tilde \chi_i$ in terms of the parameters 
$J_1, J_2, m, \tilde J_1, \tilde J_2, \tilde m$. Using that these 
parameters coincide for both solutions, one shows that the 
contribution to the boundary integral~\eqref{MazurId} vanishes. 
Again, the details of this argument only depend upon the asymptotics
of the solution, but not the horizon topology, or rod structure. 
They are therefore
identical to the arguments given in~\cite{Ida} for spherical black
holes, see also~\cite[Sec. 4.3]{Harmark}. 
\end{enumerate}
This completes the proof.
\qed

\section{Conclusion}

In this paper we have considered 5-dimensional stationary, 
asymptotically flat, vacuum black hole spacetimes with 2 commuting axial 
Killing fields generating an action of $U(1) \times U(1)$. Under the 
additional hypothesis that there are no points with a discrete
isotropy subgroup, we have shown that the black hole
must have horizon topology $S^3, S^2 \times
S^1$, or $L(p,q)$, and that each
solution is uniquely specified by the asymptotic charges (mass and the
two angular momenta), together with certain data describing the 
relative position and distance of the horizon and axis of rotation---the
``rod-structure,'' defined in a somewhat different form 
first by~\cite{Harmark}. 
Our proof mostly relied on the known
$\sigma$-model formulation of the reduced Einstein
equation~\cite{Maison,Ida}, combined with basic 
arguments clarifying the global structure of the factor manifold
of symmetry orbits. 

As we have already pointed out in the introduction, the case considered 
in this paper presumably does not represent the most general stationary, 
asymptotically flat black hole solution in 5 dimensions. It appears highly
unlikely that our method of proof could be generalized to solutions with 
only one axial Killing field, if such solutions were to exist. 
On the other hand, we believe that 
our assumption that there are no points with a discrete isotropy group
is only of a technical nature. 
Without this assumption, the orbit 
space will have singular points (``orbifold points''), rather than being 
an analytic manifold with boundary. Our analysis of the integrated 
divergence identity~\eqref{MazurIdint} 
then would also have to include the boundaries resulting from 
the blow ups of the orbifold points. 
It seems not unlikely that the proof could still go through if the 
nature of the discrete subgroups is identical 
for the two solutions. Thus, it appears that we need to specify in general 
at least
(a) the mass and angular momenta (b) rod structure, and (c) a datum 
describing the position of the points with discrete isotropy subgroups in 
the upper half plane, together with the specification of the subgroups 
themselves.

It is also interesting to ask how the parameters in the rod structure are 
related to other properties of the solution, such as the invariant
charges, horizon area, or surface gravity. For example, by 
evaluating the horizon area for the metric~\eqref{metric}, one finds
that the rod parameter $l_h$ associated with the horizon is given by 
$l_h = \kappa A/4 \pi^2$, but we do not know whether similar relations
exist for the other rod parameters. It is also not clear 
whether all rod structures can actually be realized 
in solutions to the vacuum equations, nor whether the 
horizon topologies $L(p,q)$ can be realized\footnote{
Solutions with horizon topology $L(n,1)$ have however been 
found in Einstein-Maxwell theory, see~\cite{Ishihara}.
}. Finally it would be interesting to see if the constructions of this 
paper can be generalized to include matter fields~\cite{Hollands}.

\vspace{1cm}

\noindent
{\bf Acknowledgements:} We would like to thank Troels Harmark 
for comments and Thomas Schick for discussions. S. Y. would like to 
thank the Alexander von Humboldt Foundation for a stipend, and 
the Institut f\" ur Theoretische Physik G\" ottingen for its 
kind hospitality. He also acknowledges financial support from 
the Bulgarian National Science Fund under Grant MUF04/05 (MU 408). 

\medskip
\noindent
{\bf Note added in proof:} After this manuscript was posted, it 
was noted by P. Chrusciel that our analysis did not 
properly take into account points with discrete isotropy group. 
We have added a corresponding assumption to the hypothesis. We are grateful 
to him for sharing his insight with us.


\begin{thebibliography}{99}

\bibitem{Adams}
Adams, C.C.: {\em The Knot Book: An Elementary Introduction 
to the Mathematical Theory of Knots} New York, W. H. Freeman (1994)

\bibitem{Bunting}
Bunting,~G.~L.: 
Proof of the uniqueness conjecture for black holes, 
(PhD Thesis, Univ. of New England, Armidale, N.S.W., 1983)

\bibitem{Carter}
Carter,~B.: 
Axisymmetric black hole has only two degrees of freedom,  
Phys. Rev. Lett. {\bf 26}, 331-333 (1971)

\bibitem{KK2}
Y. M. Cho and P. G. O. Freund:
Non-Abelian gauge fields as Nambu-Goldstone fields, 
Phys. Rev. D {\bf 12}, 1711 (1975)

\bibitem{Chrusciel} 
Chru\'sciel, P.T.: On rigidity of analytic black holes,  
Commun. Math. Phys. {\bf 189}, 1-7 (1997)  

\bibitem{Emparan}
Emparan,~R. and Reall,~H.~S.: A rotating black ring in five dimensions. 
Phys.\ Rev.\ Lett.\  {\bf 88}, 101101 (2002) 

\bibitem{Reall}
Emparan,~R. and Reall,~H.~S.: Generalized Weyl solutions, Phys. Rev. D
{\bf 65}, 084025 (2002)


\bibitem{Friedrich} 
Friedrich, H., Racz, I. and Wald, R.M.: 
On the rigidity theorem for spacetimes with a stationary event horizon 
or a compact Cauchy horizon,  
Commun. Math. Phys. {\bf 204}, 691-707 (1999)  

\bibitem{Woolgar}
Galloway,~G.~J., Schleich,~K.,~Witt,~D.~M., and Woolgar,~E.: 
Topological censorship and higher genus black holes, 
Phys.\ Rev.\ D {\bf 60}, 104039 (1999)

\bibitem{Woolgar1}
Galloway,~G.~J., Schleich,~K., Witt.~D., and Woolgar,~E.: 
The AdS/CFT correspondence conjecture and topological censorship, 
Phys.\ Lett.\ B {\bf 505}, 255 (2001)


\bibitem{Galloway}
Galloway, G.J., Schoen, R.: A Generalization of Hawking's 
black hole topology theorem to higher dimensions, 
Commun. Math. Phys. {\bf 266}, 571 (2006)

\bibitem{Gibbons}
Gibbons~G.~W., Ida,~D., and Shiromizu,~T.: 
Uniqueness and non-uniqueness of static black holes in higher dimensions,
Phys.\ Rev.\ Lett.\  {\bf 89}, 041101 (2002)

\bibitem{Oelsen}
  Harmark~T. and Olesen,~P:
  On the structure of stationary and axisymmetric metrics,
  Phys.\ Rev.\  D {\bf 72}, 124017 (2005)
  [arXiv:hep-th/0508208].

\bibitem{Harmark}
  Harmark, T.:
  Stationary and axisymmetric solutions of higher-dimensional general
  relativity,
  Phys.\ Rev.\  D {\bf 70}, 124002 (2004)
  [arXiv:hep-th/0408141].

\bibitem{H72}
Hawking, S.W.: 
Black holes in general relativity. 
Commun. Math. Phys. {\bf 25}, 152-166 (1972)  

\bibitem{Hawking}
Hawking, S.W. and Ellis, G.F.R.: 
{\em The large scale structure of space-time}  
Cambridge: Cambridge University Press, 1973 

\bibitem{Heusler}
Heusler, M.: {\em Black hole uniqueness theorems,}
Cambridge University Press (1996)

\bibitem{HIW}
  Hollands, S., Ishibashi, A. and Wald, R. M.:
  A higher dimensional stationary rotating black hole must be
  axisymmetric,
  Commun.\ Math.\ Phys.\  {\bf 271}, 699 (2007)
  [arXiv:gr-qc/0605106].

\bibitem{HI}
  Hollands, S. and Ishibashi, A.:
  Asymptotic flatness and Bondi energy in higher dimensional gravity,
  J.\ Math.\ Phys.\  {\bf 46}, 022503 (2005)
  [arXiv:gr-qc/0304054].

\bibitem{Hollands}
Hollands, S. and Yazadjiev, S., 
Work in progress 

\bibitem{Ishihara}
Ishihara, H., Kimura, M., Masuno, K., Tomizawa, S.:
Black holes on Euguchi-Hanson space in five-dimensional 
Einstein Maxwell theory, Phys. Rev. D {\bf 74}, 047501 (2006)

\bibitem{Israel67}
Israel,~W.: 
Event horizons in static vacuum space-times, 
Phys. Rev., {\bf 164}, 1776-1779 (1967) 

\bibitem{KK1}
R. Kerner: 
Generalization of Kaluza-Klein theory for an 
arbitrary non-abelian gauge group, 
Ann. Inst. H. Poincare, {\bf 9}, 143 (1968)

\bibitem{Kobayashi}
Kobayshi, S., Nomizu, K.: {\em Foundations of Differential 
Geometry I}, Wiley 1969

\bibitem{Maison}
Maison,~D.: Ehlers-Harrison-type Transformations for Jordan's
extended theory of graviation, Gen. Rel. Grav. {\bf 10}, 717 (1979)

\bibitem{Ida}
Morisawa, Y., Ida, D.: A boundary value problem for five-dimensional 
stationary black holes, Phys. Rev. D {\bf 69}, 124005 (2004)

\bibitem{Mazur}
Mazur,~P.~O.: 
Proof of uniqueness of the Kerr-Newman black hole solution,
J. Phys. A, {\bf 15}, 3173-3180 (1982) 

\bibitem{Moncrief}
Moncrief, V. and Isenberg, J.: 
Symmetries of cosmological Cauchy horizons. 
Commun. Math. Phys. {\bf 89}, 387-413 (1983) 

\bibitem{Myers} 
Myers, R.C. and Perry, M.J.: 
Black holes in higher dimensional space-times. 
Annals Phys. {\bf 172} 304 (1986) 

\bibitem{Racz}
Racz,~I.: 
On further generalization of the rigidity theorem for spacetimes 
with a stationary event horizon or a compact Cauchy horizon. 
Class.\ Quant.\ Grav.\  {\bf 17}, 153 (2000) 

\bibitem{Robinson} 
Robinson,~D.~C.: 
Uniqueness of the Kerr black hole, 
Phys. Rev. Lett. {\bf 34}, 905-906 (1975) 

\bibitem{Sudarsky} 
Sudarsky, D. and Wald, R.M.: 
Extrema of mass, stationarity, and staticity, and solutions to 
the Einstein Yang-Mills equations. 
Phys. Rev. D {\bf 46} 1453-1474 (1992) 

\bibitem{Waldbook} 
Wald, R.M.: {\em General Relativity}. Chicago: 
University of Chicago Press, 1984 

\bibitem{Weinstein}
Weinstein, G.: On rotating black holes in 
equilibrium in general relativity, 
Commun. Pure Appl. Math. {\bf 43}, 903 (1990)
\end{thebibliography}
\end{document}